\documentclass[12pt]{article} 
\begin{document}

\newcommand{\sect}[1]{\setcounter{equation}{0}\section{#1}}
\renewcommand{\theequation}{\thesection.\arabic{equation}}
\newcommand{\be}{\begin{equation}}
\newcommand{\ee}{\end{equation}}
\newcommand{\bea}{\begin{eqnarray}}
\newcommand{\eea}{\end{eqnarray}}
\newcommand{\beano}{\begin{eqnarray*}}
\newcommand{\eeano}{\end{eqnarray*}}
\pagestyle{empty}


\newcommand{\LAP}{LAPTH}
\def\logo{{\bf {\huge LAPTH}}}

\centerline{\logo}

\vspace {.3cm}

\centerline{{\bf{\it\Large 
Laboratoire d'Annecy-le-Vieux de Physique Th\'eorique}}}

\centerline{\rule{12cm}{.42mm}}

\vspace{1.2em}
 \addtolength{\baselineskip}{1.2ex}

\begin{center}
{\LARGE\bf Deformed oscillators algebra formulation of the Nonlinear 
Schr\" odinger 
hierarchy and of its symmetry}
\\[2.1em]
 
{\Large E. Ragoucy}\footnote{ragoucy@lapp.in2p3.fr}

\null

\noindent
{\it LAPTH\\ 
UMR 5108 du CNRS associ\'ee \`a l'Universit\'e de Savoie \\
BP 110, F-74941 Annecy-le-Vieux, France}
\end{center}

\vfill
\addtolength{\baselineskip}{-1.2ex}

\begin{abstract}
We present a self-contained formulation of the Nonlinear Schr\" odinger 
hierarchy and its Yangian symmetry in terms of deformed oscilator 
algebra (Z.F. algebra). The link between Yangian $Y(gl_{N})$ and 
finite $W(gl_{pN},N.gl_{p})$ algebras is also illustrated in this 
framework.
\end{abstract}
\vfill
\centerline{\it To be published in the proceedings of QGIS-99, Prague, 
June 17-19 1999}
\vfill
\rightline{July 1999}
\rightline{LAPTH-743/99-Conf}
\rightline{\tt hep-th/9907218}
\newpage
\pagestyle{plain}
\setcounter{page}{1}

The present text is a summary of the original work \cite{MRSZ}, done in 
collaboration with M. Mintchev, P. Sorba and Ph. Zaugg. The proofs and 
detailed calculations can be found there, as well as a more precise 
bibliography.

\section{The NLS equation}
We start with the $N$-vectorial version of the Non Linear Schr{\"o}dinger equation 
(NLS) in 1+1 dimension, on the real line:
\be
i\frac{\partial\phi(x,t)}{\partial t} +
\frac{\partial^2\phi}{\partial^2 x} = 2g\ (\phi^\dag\cdot\phi) \phi
\mbox{ with } 
\phi=\left(\begin{array}{c}\varphi_1 \\ \varphi_2 \\
\vdots \\ \varphi_N\end{array}\right)
\ee
where $\varphi_{j}$ are complex fields, and the coupling constant $g$ is 
chosen real and negative (repulsive case). The corresponding 
Hamiltonian is given by
\be
H = \int dx \left(\frac{\partial \varphi^{\dag i}}{\partial x} 
\frac{\partial \varphi_i}{\partial x} + g \varphi^{\dag i} \varphi^{\dag 
j} \varphi_i \varphi_j \right),
\ee

The solution is given by a series expansion in powers of $g$:
\bea
\phi(x,t) &=& \sum_{\ell=0}^\infty (-g)^\ell \phi^{(\ell)}(x,t) \
\mbox{ where } \phi^{(\ell)}=\left(\begin{array}{c}\varphi^{(\ell)}_1 \\ 
\varphi^{(\ell)}_2 \\
\vdots \\ \varphi^{(\ell)}_N\end{array}\right)
\label{phi:g}\\
\varphi^{(\ell)}_{j_0}(x,t) &=&\! \int 
\frac{d^{\ell+1}\nu\, d^\ell \mu}{(2\pi)^{2\ell+1}}\ 
\frac{a^{\dag j_1}(\mu_1) 
a^{\dag j_2}(\mu_2)\cdots a^{\dag j_\ell}(\mu_\ell)
a_{j_\ell}(\nu_\ell) \cdots a_{j_1}(\nu_1)a_{j_0}(\nu_0)}
{\prod_{j=1}^\ell(\mu_j-\nu_j-i\epsilon)(\mu_j-\nu_{j-1}-i\epsilon)}\ 
e^{i\Omega_\ell}
\nonumber\\
\Omega_\ell &=& \sum_{j=0}^\ell (x\nu_j-t\nu^2_j)-\sum_{j=1}^\ell (x\mu_j-t\mu^2_j)
\eea
This solution is valid for the classical case \cite{Ros} ($a$'s are 
then arbitrary functions) as well as for the quantum case 
\cite{solQ}, where in that case $a(\mu)$ and $a^\dag(\mu)$ generate a deformed oscillator algebra
(or ZF algebra \cite{ZF}):
\bea
a_i(\mu)a_j(\nu) &=& R_{ji}^{lk}(\nu-\mu)
a_k(\nu)a_l(\mu) \label{aa}\\
{a^\dag}^i(\mu){a^\dag}^j(\nu) &=&
{a^\dag}^k(\nu){a^\dag}^l(\mu)
R^{ij}_{kl}(\nu-\mu)\\
a_i(\mu){a^\dag}^j(\nu) &=& 
{a^\dag}^k(\nu)R_{ik}^{jl}(\mu-\nu)
a_l(\mu)+\delta_i^j\delta(\mu-\nu)
\label{a*a}
\eea
Note that the indices $i$, $j,\dots$ run from 1
to $N$, because the vector $\phi$ is in the fundamental
representation of $sl(N)$. In the quantum case, the constructed fields 
obey canonical commutation relations:
\be\begin{array}{ll}
{[\varphi_{j}(x,t), \varphi_{k}(y,t)]} = 0 & ;\ 
{[\varphi^{\dag j}(x,t), \varphi^{\dag k}(y,t)]}=0 \\
{[\varphi_{j}(x,t), \varphi^{\dag k}(y,t)]} = 
\delta^k_{j}\delta(x-y) &
\end{array}\ee
The R-matrix appearing in this algebra is just the one of 
$Y(sl(N))$.
This is not surprising, since $Y(sl(N))$ is a symmetry of NLS. In
fact, it has been shown in \cite{Wad} that, for $N=2$, the 
Yangian generators
can be expressed in term of $\phi$ and Pauli matrices $t^a$:
\bea
J^a &=& \int dx\, \phi^\dag(x)t^a\phi(x)\label{Q:phi}\\
S^a &=& \! \frac{i}{2}\int dx\, \phi^\dag(x)t^a\partial_x\phi(x)
-\frac{ig}{2}\int dxdy\ sgn(y-x)
\left(\phi^\dag(x)t^a\phi(y)\right)
\phi^\dag(x)\cdot\phi(y)\nonumber
\eea
where $J^a$ and $S^a$ stand for the Yangian generators $Q^a_0$ and 
 $Q^a_1$ in the Drinfel'd presentation of the Yangians \cite{drin}.
 
\section{The NLS hierarchy}
The Yangian symmetry of the NSL equation (and in fact of its whole 
hierarchy) is nicely formulated using the Quantum Inverse Scattering 
Method (QISM).

We start with the Lax operator of the QISM 
\be
L(x|\lambda) = i \frac{\lambda}{2} \Sigma + \Omega(x), 
\quad\hbox{with}\quad 
\Sigma = \left(\begin{array}{ccccc} 1 & 0 & 0 & \ldots & 0 \\
0 & 1 & 0 & \ldots & 0 \\
\vdots & \ddots & \ddots & \ddots & \vdots \\
0 & \ldots & 0 & 1 & 0 \\
0 & \ldots &  & 0 & -1\end{array}\right),
\ee
\be
\hbox{and}\quad\Omega(x)= i \sqrt{g} 
\left(\begin{array}{cc}\mbox{\LARGE 0} & 
\begin{array}{c}\varphi_1(x) \\ \varphi_2(x) \\ \vdots \\ \varphi_N(x) 
    \end{array} \\
\begin{array}{cccc} - \varphi^{\dag 1}(x) & - \varphi^{\dag 2}(x) & \ldots
    & - \varphi^{\dag N}(x) \end{array} & 0
\end{array}\right)
\ee
and look at the quantum monodromy matrix $T(x,y|\lambda)$, defined by the 
equations
\be
\frac{\partial}{\partial x} T(x,y|\lambda) = \ :\! L(x|\lambda) 
T(x,y|\lambda)\! :,
\qquad
T(x,y|\lambda)\vert_{x=y}=I_{N+1},
\label{dTLT}
\ee
where $I_{N+1}$ is the identity matrix.
The infinite volume monodromy matrix
$T(\lambda)$ is then formally defined by
$$
T(\lambda) = \lim_{x \to \infty,y \to -\infty} E(-x|\lambda) 
T(x,y|\lambda) E(y|\lambda),
\quad\hbox{where}\quad  E(x|\lambda)=\exp(i \lambda x \Sigma/2).
$$
The commutation relations of $T(\lambda)$ are computed using the 
canonical commutators of the $\varphi$'s and are encoded in the 
exchange relation
\be
R^+(\lambda-\mu)\ T(\lambda) \otimes T(\mu) =
T(\mu) \otimes T(\lambda)\
R^-(\lambda-\mu)
\label{RTT}
\ee
with the $R$-matrices
\bea
 R^\pm(\mu) &=& \frac{-i g}{\mu-i g}\frac{1}{\mu} \ 
E_{jj}\otimes E_{kk}+ \frac{1}{\mu-i g}\ E_{\alpha j}\otimes 
E_{j\alpha} \nonumber\\
 &+&\frac{\mu+i g}{(\mu+i0)^2}\ E_{j,N+1}\otimes E_{N+1,j} 
+\frac{1}{\mu}\ 
E_{N+1,N+1}\otimes E_{N+1,N+1}\nonumber\\
 &\pm& i\pi\delta(\mu)(E_{jj}\otimes E_{N+1,N+1}
-E_{N+1,N+1}\otimes E_{jj}),
\eea
where the latin (resp. greek) indices run from 1 to $N$ (resp. 
$N+1$), and $E_{ab}$ is the $(N+1)\times(N+1)$ matrix with 1 at position $(a,b)$.
The term $i0$ is a consequence of the principal value regularisation 
adopted when $\mu$ goes to zero.

It will be convenient to rename some elements of the monodromy matrix 
such as $D(\lambda)=T_{N+1,N+1}(\lambda)$ and 
$b^j(\lambda)=T_{N+1,j}(\lambda)$.  Further examination of some 
components of (\ref{RTT}) yields the following relations
\bea
D(\lambda) D(\mu) &=&  D(\mu) D(\lambda), \label{DD} \\
D(\lambda) b^j(\mu) &=& \frac{\lambda-\mu+ig}{\lambda-\mu+i0}\ 
b^j(\mu) D(\lambda), \label{Db}\\
b^j(\lambda) b^k(\mu)& =&
\frac{\lambda-\mu}{\lambda-\mu-ig}\ b^k(\mu)b^j(\lambda)-
\frac{ig}{\lambda-\mu-ig}\ b^j(\mu)b^k(\lambda).
\label{bb}
\eea
The matrix element $D(\lambda)$ serves as a generating 
operator-function for the commuting integrals of motion of the NLS 
model: $D(\lambda)=1+\sum_{n=0}^\infty H_{(n)} \lambda^{-n-1}$.
Consequently, eq.~(\ref{DD}) implies that the $H_{(n)}$'s
are all in involution. Indeed, one can compute that 
$H_{(0)}$ is proportional to $\int dx \ \phi^{\dag j} \phi_j$ 
(particle number operator), 
$H_{(1)}$ to (up to $H_{(0)}$ terms) 
$- i \int dx \ \phi^{\dag j} \partial\phi_j$ (momentum) and $H_{(2)}$ 
is the NLS Hamiltonian. The other $H_{(n)}$'s ($n>2$) define higher 
Hamiltonians which define the NLS hierarchy. The operators 
$b^j(\mu)$ just correspond to the creation operators of the ZF algebra 
previously mentionned. More precisely, one has 
$a^{\dag j}(\lambda) = \frac{i}{\sqrt{g}} b^j(\lambda) 
D^{-1}(\lambda)$, and the commutation relations (\ref{Db}-\ref{bb}) 
ensure the right exchange relation for the $a^{\dag j}(\lambda)$'s. 
The operators $a_{j}(\lambda)$ are the adjoint operators of these 
$a^{\dag j}(\lambda)$'s.

Finally, considering $\tilde{T}(\lambda)$, the $N\times N$ submatrix   
of $T(\lambda)$, and examining the appropriate 
components of (\ref{RTT}), one deduces the following relations
\be
\widetilde{R}(\lambda-\mu) \ \tilde{T}(\lambda) \otimes \tilde{T}(\mu) =
 \tilde{T}(\mu) \otimes \tilde{T}(\lambda) \ \widetilde{R}(\lambda-\mu)
\label{RTtTt}
\ee
with yet another $R$-matrix
\be
\widetilde{R}(\lambda-\mu) = 
(\lambda-\mu) E_{jk} \otimes E_{kj} -i g I_N \otimes I_N  .
\label{Rtilde}
\ee
This coincides precisely with the defining relation of the Yangian 
$Y(gl_{N})$.

The fact that the Yangian algebra commutes with the Hamiltonians of 
the NLS hierarchy is a consequence of the exchange relation as well, 
since one extracts from (\ref{RTT}) that
\be
[\tilde{T}_{ij}(\lambda),D(\mu)] = 0
\label{TDzero}
\ee

\section{Z.F. formulation}
The two previous sections have both made appear the ZF algebra: the 
former as a "building bock" for the quantum solutions of the NLS 
equation, and the latter as the "remains" of the matrix $T(\lambda)$ 
when one has picked up the Yangian $\tilde{T}(\lambda)$ and the 
Hamiltonians $D(\lambda)$. It is thus natural to wonder whether this 
ZF algebra can be the central element of the NLS hierarchy and of its 
symmetry.
Indeed, since the field $\phi$ is built on it and as the Yangian 
generators are constructed on $\phi$ (at least for $N=2$), one could 
think that the task is not hard. However, $\phi$ is given as a series 
expansion in $a$'s and $a^\dag$'s, and the Yangian generators are 
polynomial in $\phi$, so 
that a direct calculation is almost impossible. Alternatively, we 
 construct operators which have the right commutation relations 
with the $a$'s and $a^\dag$'s: since the Fock space spanned by these 
latter is dense in the Hilbert space of the NLS model, this will be 
enough to identify the operators. 
\subsection{Hamiltonians of the hierarchy}
We introduce the operators
\be
\tilde{H}_{(n)}=\int d\mu\ \mu^n a^{\dag j}(\mu) a_{j}(\mu)\label{Ha}
\ee
Using the relations (\ref{aa}-\ref{a*a}), it is easy to show that 
\be
{[\tilde{H}_{(n)}, a^{\dag j}(\mu)]} = \mu^n a^{\dag j}(\mu) 
\quad ; \quad 
{[\tilde{H}_{(n)}, a_{j}(\mu)]} = -\mu^n a_{ j}(\mu) \label{comHa}
\ee
which is just the definition of the Hamiltonians in the quantum NLS 
hierarchy. 

Thanks to the simple expression (\ref{Ha}), it is easy to deduce the 
general solution to the NLS hierarchy. In fact, the local fields 
$\phi$ which has a time evolution given by the Hamiltonian 
${H}_{(n)}\sim\tilde{H}_{(n)}$ have the same expansion 
(\ref{phi:g}) with now
\be
\Omega^{(n)}_\ell = \sum_{j=0}^\ell (x\nu_j-t\nu^{n}_j)-\sum_{j=1}^\ell 
(x\mu_j-t\mu^n_j)
\ee
\subsection{Yangian generators}
In the same way, we define the operators
\be 
J^a = \sum_{\ell=1}^\infty \frac{(-)^{\ell+1}}{\ell!} J^a_{(\ell)} 
\quad\hbox{and}\quad
S^a = \sum_{\ell=1}^\infty \frac{(-)^{\ell+1}}{\ell!} S^a_{(\ell)} 
\ee
with
\bea
\!\! J^a_{(\ell)} &=&\!\!  \int d^\ell \mu\ a^{\dag j_1}(\mu_1)a^{\dag j_2}(\mu_2)
\,.. a^{\dag j_\ell}(\mu_\ell) \ 
(T^a)_{j_1j_2\dots j_\ell}^{k_1k_2\dots k_\ell}\ 
a_{k_\ell}(\mu_\ell)\,.. 
a_{k_1}(\mu_1)\label{defJn}\\
\!\! S^a_{(\ell)} &=&\!\! \int d^\ell \mu\ a^{\dag j_1}(\mu_1)
a^{\dag j_2}(\mu_2)\,..  
a^{\dag j_\ell}(\mu_\ell) \ 
(\tilde{T}^a)_{j_1j_2\dots j_\ell}^{k_1k_2\dots k_\ell}\ a_{k_\ell}(\mu_\ell)
\,..  a_{k_1}(\mu_1)
\label{defSn}
\eea
The tensor $T^a$ is given by
\be
(T^a)_{j_1j_2\dots j_\ell}^{k_1k_2\dots k_\ell}= \sum_{m=1}^\ell (-)^{m-1}
{\ell-1\choose m-1}\ 
(t^a)^{k_{m}}_{j_{m}}\ \delta_{j_{1}}^{k_{1}}\ldots
\delta_{j_{m-1}}^{k_{m-1}}\delta_{j_{m+1}}^{k_{m+1}}\ldots 
\delta_{j_{\ell}}^{k_{\ell}}.
\label{defT}
\ee
where $t^a$ are the generators of $sl_N$ in the fundamental 
representation ($[t^a,t^b]=i{f^{ab}}_c\, t^c$).
With this form, $J^a$ satisfies
\bea
 {[J^a,a^{\dag k}(\mu)]} &=& (a^{\dag}(\mu) t^a)^k, \quad
 {[J^a,a_{k}(\mu)]} \ =\ -(t^aa(\mu) )_k \\
 {[J^a,J^b]} &=& i{f^{ab}}_{c}\, J^c
\eea
Similarly, omitting the obvious $\delta^{j_{r}}_{k_{r}}$ symbols, we 
write 
\be
(\tilde{T}^a)_{j_1j_2\dots j_\ell}^{k_1k_2\dots k_\ell}= 
\sum_{m=1}^\ell (-)^{m-1}{\ell-1\choose m-1}\ \left\{
\mu_m\, (t^a)^{k_{m}}_{j_{m}}\, -\frac{g}{2}{f^{a}}_{bc}\, 
\sum_{i=1}^{m-1}  (t^{b})^{k_{i}}_{j_{i}} (t^c)^{k_{m}}_{j_{m}}\right\}.
\ee
With this definition, one gets
\bea
 {[S^a,a^{\dag k}(\mu)]} &=& \mu\, (a^{\dag}(\mu)\, 
t^a)^k-\frac{g}{2}{f^a}_{bc} (a^{\dag}(\mu)\, t^c)^k J^b, \\
 {[S^a,a_{k}(\mu)]} &=& -\mu\, (t^a\,a(\mu) )_k-
 \frac{g}{2}{f^a}_{bc} J^b(t^c\,a(\mu) )_k , \\
  {[J^a,S^b]} &=& i{f^{ab}}_{c}\, S^c
\eea
which are the required commutators for the Yangian 
generators $Q_0^a$ and $Q_1^a$.

Note that the Yangian generators are defined through a series 
expansion, not in the coupling constant, but rather in the number of 
creation operators $a^\dag$. In the case of the $\phi$ fields, these 
two types of series indeed coincide.

Let us also remark that the form (\ref{defJn}-\ref{defSn}) clearly 
shows (using the relations (\ref{comHa})) that $J^a$ and 
$S^a$ commute with the Hamiltonians ${H}_{(n)}$, so that the Yangian 
is a manifest symmetry of the whole NLS hierarchy.

\section{Fock space and finite $W(gl_{pN},N.gl_{p})$ algebras}
It is now known \cite{13} that there is an algebra homomorphism 
between the Yangian $Y(gl_N)$ and the finite $W(gl_{pN},N.gl_{p})$ 
algebras (for any $p$). This links can be illustrated in the framework 
of NLS hierarchy, using the ZF algebra.

Indeed, if one considers the Fock space $\cal F$ spanned by the $a^\dag$'s, it 
is easy to see that it is built on subspaces ${\cal 
F}_{p}$ with fixed particle number (ie 
number of $a^\dag$). Since the Yangian generators commute with the 
particle number operator, one can consider their restriction to ${\cal 
F}_{p}$. In that case, the series expansion defining the Yangian 
generators troncates at level $p$, and we are left with a polynomial 
(finite dimensional) algebra. This algebra is nothing but the 
$W(gl_{pN},N.gl_{p})$ algebra (in a special representation). 

Thus, while on the full space $\cal F$ we have the complete Yangian 
symmetry, on each subspace ${\cal F}_{p}$ the Yangian troncates to a 
finite $W(gl_{pN},N.gl_{p})$ algebra which leaves ${\cal F}_{p}$
 (globally) invariant.


\begin{thebibliography}{99}
\bibitem{MRSZ} M. Mintchev, E. Ragoucy, P. Sorba, Ph. Zaugg, {\tt 
hep-th/9905105}, to be published in J. Phys. {\bf A} (1999);\\ See also
{\tt hep-th/9812186} or proceedings of {\it
  Yang-Baxter systems, non linear models and their 
applications}, Seoul (Korea) October 20-23, 1998.

\bibitem{Ros} R. Rosales, Stud. Appl. Math. {\bf 59} (1978) 117.

\bibitem{solQ}  E. Sklyanin and L. D. Faddeev, Sov. Phys. Dokl.
   {\bf 23} (1978) 902; \\
E. Sklyanin, Sov. Phys. Dokl. {\bf 24} (1979) 107;\\
H.B. Tacker, D. Wilkinson, Phys. Rev. {\bf D19} (1979) 3660;\\
 D.B. Creamer, H.B. Tacker, D. Wilkinson, Phys. Rev. {\bf D21} (1980) 1523;\\
 J. Honerkamp, P. Weber, A. Wiesler, Nucl. Phys. {\bf B152} (1979) 266;\\
 B. Davies, J. Phys. {\bf A14} (1981) 2631.

\bibitem{ZF} A.B. Zamolodchikov, A.B. Zamolodchikov, Ann. Phys. {\bf
    120} (1979) 253; \\
    L.D. Faddeev, Sov. Scient. Rev. {\bf C1} (1980) 107.

\bibitem{Wad} S. Murakami, M. Wadati, J. Phys. {\bf A29} (1996)
  7903.

\bibitem{drin} V.G. Drinfel'd, Sov. Math. Dokl. {\bf 32} (1985) 254.

\bibitem{13} E. Ragoucy, P. Sorba, Comm. Math. Phys. {\bf 203} (1999) 
551;\\
See also Cz. J. Phys. {\bf 48} (1998) 1483, or
{\tt hep-th/9803242}.

\end{thebibliography}
\end{document}